\documentclass{ws-procs9x6}
\usepackage{graphicx}
\begin{document}

\title{Bose-Einstein condensates on a permanent magnetic film atom
chip}

\author{B.~V. HALL, S. WHITLOCK, F. SCHARNBERG\footnote{\uppercase{P}resent address:
\uppercase{I}nstit\"{u}t f\"{u}r \uppercase{Q}uantenoptik,
\uppercase{U}niversit\"{a}t \uppercase{H}annover, 30167
\uppercase{H}annover, \uppercase{G}ermany}, P.~
HANNAFORD~and~A.~SIDOROV}

\address{ARC Centre of Excellence for Quantum-Atom Optics and \\Centre for Atom Optics and Ultrafast Spectroscopy \\
Swinburne University of Technology, Melbourne, Victoria 3122, Australia\\
E-mail: brhall@swin.edu.au}

\maketitle \abstracts{We present a permanent magnetic film atom chip
based on perpendicularly magnetized TbGdFeCo films. This chip
routinely produces a Bose-Einstein condensate (BEC) of 10$^5$
$^{87}$Rb atoms using the magnetic film potential. Fragmentation
observed near the film surface provides unique opportunities to
study BEC in a disordered potential. We show this potential can be
used to simultaneously produce multiple spatially separated
condensates. We exploit part of this potential to realize a
time-dependent double well system for splitting a condensate.}

\section{Introduction}
A recent advance in the field of quantum degenerate gases has been
the development of the atom chip\cite{Han01,Ott01,Fol02}. This
device streamlines BEC production by exploiting tightly confining
atom traps realized near a micro-structured magnetic field
source\cite{Du04}.  Additionally these sources can be tailored for
the manipulation of coherent matter waves and recently used to
demonstrate interference of a BEC in a double well
potential\cite{Shi05,Sch05a}. Several technical limitations have
been identified for atom chips where micro-structured
current-carrying wires provide magnetic confinement\cite{Hen03}.
Current noise can cause spin flip loss in addition to in-trap
heating. Methods of conductor fabrication also impose limitations as
wire edge roughness induces fragmentation of ultracold atom
clouds\cite{Est04}. Moreover a fundamental spin-flip loss mechanism
caused by thermally induced current fluctuations is inherently large
in conducting materials\cite{Jon03}.
 Some of these limitations can be addressed though the implementation
of atom chips which use permanent magnetic materials as configurable
magnetic field sources\cite{Hin99,Eri04}. Permanent magnetic
materials have intrinsically low magnetic field noise, exhibit
reduced spin-flip loss rates and allow tighter confinement close to
the magnet surface\cite{Sin05b,Sch05b}.

In this paper we report on recent experiments performed with an atom
chip that incorporates a perpendicularly magnetized, permanent\
magnetic film.

\section{Atom chip: design and construction}

\begin{figure}[htb]
\centerline{\includegraphics{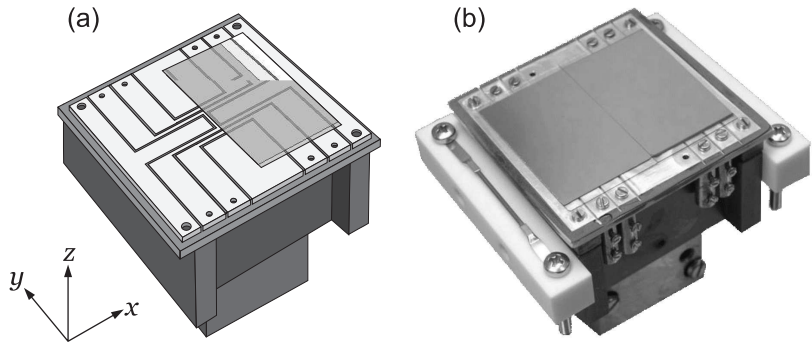}}
\caption{\label{chipfig}A schematic (a) and an image (b) of the atom
chip. From the top down: two adjacent gold coated glass slides
(1$\times$TbGdFeCo, 1$\times$blank), machined Ag foil H-shape wire
with end wires, Shapal-M base-plate and Cu heat sink.}
\end{figure}

A schematic diagram and photographic image of the permanent magnetic
film atom chip is shown in Fig.~\ref{chipfig}. The atom chip is
based upon a single rectangular slab of uniformly magnetized
TbGdFeCo film with a thickness $h$.  The film is mounted in the $xy$
plane, with one long edge parallel to the $y$~axis . The magnetic
field produced by this slab is the same as that from an equivalent
current ($I_{eff}=hM_R$) that propagates around the slab
perimeter\cite{Jac99}. The magnitude of the field near this edge is
given by
\begin{equation}
\label{bfilm} B_{film}=\frac{\mu_0}{2 \pi}\frac{hM_R}{z},
\end{equation}
where $z$ is the distance from the film.  A uniform field $B_{x}$
cancels the film field at a height $z_0$, resulting in a
two-dimensional quadrupole potential which can be used as an atomic
waveguide for weak-field seeking atoms.  A three-dimensional (3D)
magnetic trap is achieved by pinching off this guide with a
nonuniform axial field $B_y$. This field is made by two parallel
current-carrying end wires, located beneath and perpendicular to the
waveguide.  A nonzero $B_y$ suppresses Majorana spin-flip loss while
creating a harmonic potential with a radial frequency
\begin{equation}
\label{frad} 2 \pi f_{rad} =
\frac{\mu_0}{2\pi}\frac{hM_R}{{z_0}^2}\sqrt{\frac{\mu_Bg_Fm_F}{mB_y}},
\end{equation}
where  $\mu_Bg_Fm_F$ is the usual Zeeman factor and $m$ is the
atomic mass.

The top layer of the atom chip consists of two adjacent
40$\times$23$\times$0.3~mm$^3$ glass slides.  One slide has a
uniformly magnetized TbGdFeCo coating and both have a reflective
gold overlayer. The large reflective area is beneficial for
efficiently collecting atoms in a reflection magneto-optical trap
(MOT) near the surface within a single UHV system. A multilayered
deposition technique was used to produce the film with a composition
of Tb$_6$Gd$_{9.6}$Fe$_{80}$Co$_{4.4}$. This material has properties
advantageous to integrated atom optics, including large
perpendicular anisotropy, high Curie temperature
(T$_C$~=~300~$^\circ$C), large coercivity ($H_C$), strong remanent
magnetization ($M_R$) and excellent magnetic homogeneity. A detailed
account of the TbGdFeCo films and the atom chip will be reported
elsewhere\cite{Hall05a}. The film used on the chip has a remanent
magnetization of $\mu_0M_R$~=~0.28~T for a total magnet thickness of
900~nm ($hM_R~=~0.20\pm0.01$~A).  A tight trap of frequency
$f_{rad}\approx~$1~kHz is obtained approximately 70~$\mu$m above the
film using modest values of $B_{x}$~=~0.56~mT and $B_y$~=~0.1~mT.
While this trap is excellent for evaporative cooling, it has neither
the depth nor the spatial extent to be loaded directly by a MOT.

To provide large, time-dependent potentials, a machined silver foil
structure was designed to sit beneath the magnetic film\cite{Val04}.
The foil is 0.5~mm thick and machined in an H-shape with additional
end wires either side (Fig.~\ref{chipfig}a). The foil is epoxied to
a Shapal-M ceramic plate and 500~$\mu$m insulating grooves were cut
with a computer controlled PCB mill. The conductor cross section is
1.0~$\times$~0.5~mm$^2$ facilitating the use of high currents
($>30~$A). The H-shape structure allows rapid switching between a
U-shape ($I_U$) or Z-shape ($I_Z$) current path. This provides a 3D
quadrupole field geometry to form a U-wire MOT or an Ioffe-Pritchard
(\textit{IP}) magnetic potential, respectively\cite{Rei99}.  The two
glass slides, silver foil structure and ceramic plate are screwed to
a copper heat sink with two rubidium dispensers and is mounted
upside down in the vacuum chamber.

\section{BEC on the permanent magnetic film}

\begin{figure}[htb]
\centerline{\includegraphics{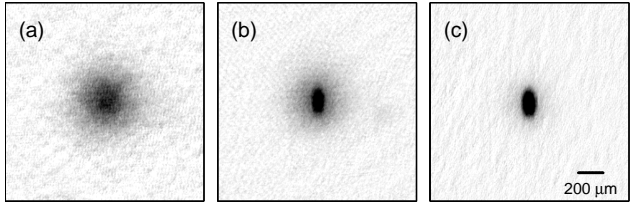}}
\caption{\label{filmbec}Absorption images of an atom cloud after
30~ms of ballistic expansion from a permanent magnetic film trap.
Images correspond to (a) a thermal cloud, (b) a partially condensed
cloud and (c) an almost pure condensate.}
\end{figure}

A reflection MOT is loaded via a pulsed Rb dispenser where
2$\times$10$^8$ atoms are collected 4.6~mm from the surface. After
15~s hold time the atoms are transferred and compressed into a
U-wire MOT located 1.6~mm from the film surface. After short
polarization gradient cooling and optical pumping stages
4$\times$10$^7$ atoms are transferred into the Z-wire $\textit{IP}$
trap.  Atoms are then moved to 560~$\mu$m from the surface where the
radial trap frequency is $2\pi\times530$~Hz and the elastic
collision rate is high enough to begin forced evaporative cooling.
Evaporation to the BEC transition is a two step process. For the
first 8.85~s of a single 10~s logarithmic radio frequency (RF)
sweep, atoms are cooled to a temperature of $\sim~5~\mu$K. Then the
RF amplitude is reduced to zero for 150~ms while the atoms are
transferred to the magnetic film trap at $z~=~90~\mu$m
($f_{rad}~=~700~$Hz, $f_{ax}~=~$8~Hz, $I_Z$~=~0~A and
$I_{end}$~=~6~A). An additional uniform magnetic field tunes the
trap bottom to minimize any discontinuity in the evaporation
trajectory. Finally the RF amplitude is increased and evaporation
continues for 1~s to the BEC phase transition. Before imaging, the
magnetic film trap is moved 170~$\mu$m from the surface. Both
$I_{end}$ and $B_{x}$ are rapidly turned off, leaving the atoms to
ballistically expand with minor acceleration from the permanent
field gradient. Figure~\ref{filmbec} shows absorption images of a
ballistically expanded cloud of atoms crossing the BEC phase
transition. The apparatus creates a new condensate of
$1\times10^5$~atoms in the magnetic film trap every 50~s.

\begin{figure}[htb]
\centerline{\includegraphics{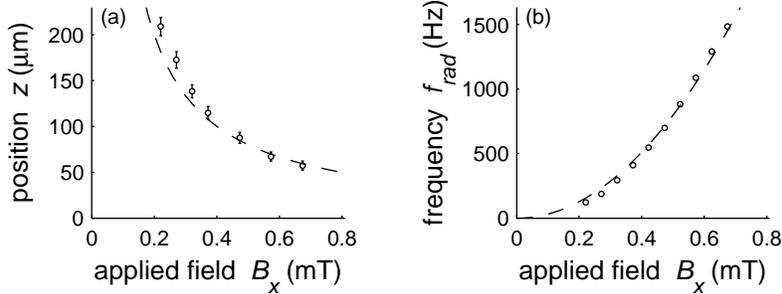}}
\caption{\label{ieff}Measurements of the trap position from the film
surface (a) and of the trap frequency (b) as a function of applied
field $B_x$.  The data (open circles) agrees well with predictions
(dashed line) of equations 1 and 2.  The discrepancy of position for
low fields can be accounted for by including the effect of gravity.}
\end{figure}

\section{Characterization of the magnetic film potential}

An {\it in situ} measurement of the film magnetization allows the
BEC to be used as a highly sensitive magnetic field probe. The small
spatial extent of the BEC allows accurate position information to be
obtained by absorption imaging. Since the BEC is localized at the
magnetic field minimum (where $B_{film} = -B_{x}$) a measure of
distance from the film versus $B_{x}$ yields $\textit{hM$_R$}$. The
narrow momentum distribution of a BEC also allows excellent
resolution when performing RF spectroscopy on the trapped atoms. A
short RF pulse provides an absolute measurement of $B_x$ by
observing the resonant population of the $F=2,m_F=-2,-1,0,1,2$
states. Each magnetic state of a freely falling cloud is spatially
separated by a field gradient provided by $I_Z$. The magnetic
potential is further characterized by observing centre of mass
oscillations.  A magnetically trapped condensate is analogous to a
high Q resonator since the trap frequency is much higher than the
damping rate\cite{Mcg04}.  By exciting radial centre of mass motion
$f_{rad}$ can be measured to better than 1~Hz.  The trap position
and the trap frequency have been measured against $B_x$ and are
shown in Fig ~\ref{ieff}. The data are consistent with the
predictions of equations 1 and 2.

In addition to the expected magnetic field from the film we observe
a small spatially alternating component of magnetic field parallel
to the film edge, $\Delta B_y$. This oscillating magnetic field
leads to fragmentation of cold clouds near the film surface and can
be compared to fragmentation observed near microfabricated
wires\cite{Est04}. Initial observations suggest that $\Delta B_y$
has an RMS amplitude of $\sim4.7~\mu$T, $90~\mu$m from the film
surface.  It appears roughly periodic with a period of $450~\mu$m,
however finer structure exists closer to the surface. The relative
amplitude of the fragmenting field $\Delta B_{y}/\Delta
B_{film}~\sim~10^{-2}$ is two orders of magnitude larger than that
observed above wire based atom chips.

\section{A string of BECs in a disordered potential}

\begin{figure}[htb]
\centerline{\includegraphics{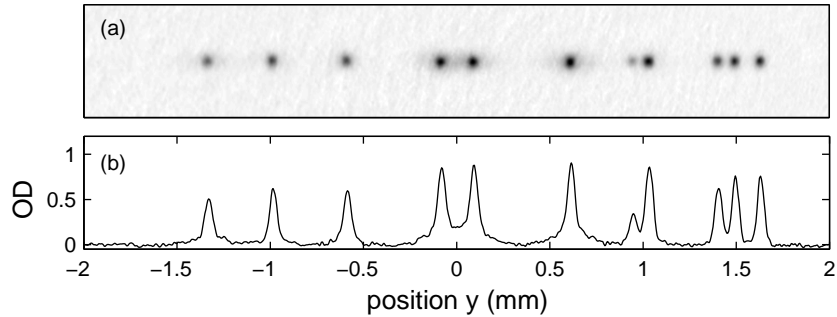}}
\caption{\label{frag}Absorption image (a) and associated optical
density profile (b) for multiple condensates realized in the
disordered potential.  Images are taken after 15~ms of ballistic
expansion.}
\end{figure}

Although the fragmented potential limits the ability to form a
smooth atomic waveguide, the micro-structure provides a unique
possibility for studying BEC in a disordered potential
(Fig.~\ref{frag}). After the initial stages of RF cooling, the
thermal cloud is transferred to the magnetic film and allowed to
expand axially under weak confinement ($f_{ax}\approx1$~Hz). The
elongated cloud spans most of the image area ($\sim5~$mm).  The
cloud is then radially compressed and moved closer to the surface
where fragmentation becomes significant.  An additional 4~s of
evaporative cooling is applied using a second RF source. The BEC
transition is reached simultaneously across 11 independent potential
wells spanning 3~mm. The total number of condensed atoms is
comparable to that observed for a single BEC
($N_{BEC}~\sim1\times10^{5}$). The large separation between
individual condensates permits addressability by optical means or by
RF coupling in a field gradient.

\section{A double well system for splitting a BEC}

\begin{figure}[htb]
\centerline{\includegraphics{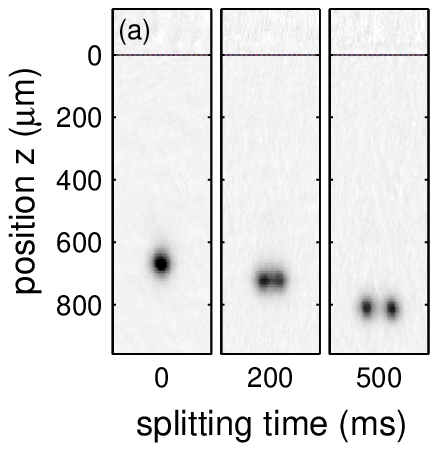}
\quad\quad\includegraphics{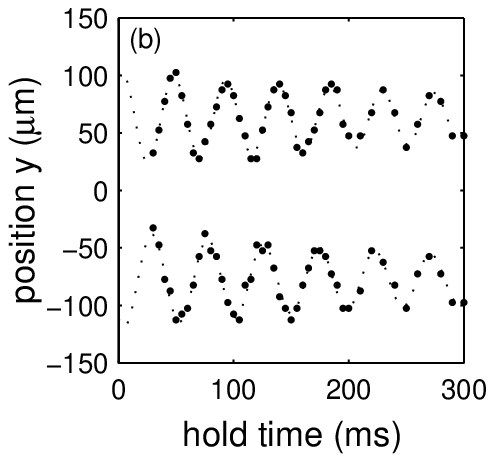}}

\caption{\label{manip}Absorption images (a) of condensates during
time dependent splitting in the double well potential. Images were
taken after 2~ms of ballistic expansion.  The relative position of
each cloud is inverted due to additional acceleration from the
permanent magnetic field gradient. Axial trap frequencies in the
double well (b) were determined by observing centre of mass
oscillations.}
\end{figure}

The disordered potential produced by the film is rich in structure.
It is possible to transport atoms to particular regions along the
film in order to perform various experiments.  A roughly symmetric
double well potential is present near the centre of Fig.~\ref{frag}.
A single BEC is formed 170~$\mu$m from the surface where the
fragmentation is small compared to the chemical potential. By
increasing $B_{x}$ the condensate moves closer to the surface and
begins to separate into two parts (Fig.~\ref{manip}a). To split the
condensate evenly, the asymmetry between the two split wells was
tuned with a magnetic field gradient to less than 100~nT.  The BEC
is separated by $\sim140~\mu$m over 500~ms to avoid unwanted
excitations.  Shorter splitting times (down to 30~ms) are possible;
however out of phase centre of mass oscillations are observed
(Fig.~\ref{frag}b).  The level of symmetry between the individual
trap frequencies is remarkable ($f_{L,rad}~=~281$~Hz,
$f_{R,rad}~=~283$~Hz, $f_{L,ax}~=~21$~Hz, $f_{R,ax}~=~$22~Hz).
Unfortunately, the resolution of our imaging system does not permit
the observation of interference fringes that would be present in
ballistic expansion. Other regions of the potential with smaller
features may provide conditions more favourable for investigating
coherent splitting and interference of BEC in a double well system.

\section{Conclusion}

We have produced an atom chip that exploits a thick perpendicularly
magnetized TbGdFeCo film for the production and manipulation of a
BEC.  We have used the BEC as a sensitive probe to directly measure
the magnetic field associated with the magnetic film.  We have shown
that this potential is rich in structure and can be exploited for
the simultaneous production of multiple condensates.  Moreover we
have used a part of this potential to demonstrate a controllable
double well system that permits the symmetric splitting of a
condensate without excitations or excessive heating.

\section*{Acknowledgments}
We thank J. Wang and D. Gough for carrying out the magnetic film
deposition.  This project is supported by the ARC Centre of
Excellence for Quantum-Atom Optics and a Swinburne University
Strategic Initiative grant.

\end{document}